\documentclass[conference,letterpaper]{IEEEtran}

\IEEEoverridecommandlockouts

\usepackage[cmex10]{amsmath}
\usepackage{amssymb}

\usepackage{cases}
\usepackage{multirow}
\usepackage{empheq}

\usepackage[noadjust]{cite}
\usepackage{verbatim}

\usepackage{algorithm}
\usepackage{algpseudocode}

\usepackage{siunitx}

\ifCLASSINFOpdf
\usepackage[pdftex]{graphicx}
\else
\usepackage[dvips]{graphicx}
\fi
\usepackage[caption=false,font=footnotesize]{subfig}

\usepackage{color}
\usepackage{multirow}

\usepackage[hyphens]{url}
\newtheorem{theorem}{Theorem}
\newtheorem{lemma}{Lemma}

\newtheorem{corollary}{Corollary}

\newtheorem{remark}{Remark}

\newtheorem{example}{Example}

\begin{document}

\title{Mutual Information Minimization for Side-Channel Attack Resistance via Optimal Noise Injection}
\author{\IEEEauthorblockN{Jiheon Woo\IEEEauthorrefmark{1},
	Donggyun Ryu\IEEEauthorrefmark{1},		
	Daewon Seo\IEEEauthorrefmark{2}, 
    Young-Sik Kim\IEEEauthorrefmark{2},
    \\Namyoon Lee\IEEEauthorrefmark{1},
	Yuval Cassuto\IEEEauthorrefmark{3}, and
	Yongjune Kim\IEEEauthorrefmark{1}}
	
	\IEEEauthorblockA{\IEEEauthorrefmark{1}Department of Electrical Engineering, POSTECH, Pohang, South Korea\\ Email: \{jhwoo1997, dgryu, nylee, yongjune\}@postech.ac.kr}
	\IEEEauthorblockA{\IEEEauthorrefmark{2}Department of Electrical Engineering and Computer Science, DGIST, Daegu, South Korea\\ Email: \{dwseo, ysk\}@dgist.ac.kr}
	\IEEEauthorblockA{\IEEEauthorrefmark{3}Viterbi Department of Electrical and Computer Engineering, Technion -- Israel Institute of Technology, Haifa, Israel \\
		Email: ycassuto@ee.technion.ac.il}

\thanks{
This work was supported by Institute of Information \& Communications Technology Planning \& Evaluation (IITP) grant funded by the Korea government (MSIT) (RS-2024-00399401, Development of Quantum-Safe Infrastructure Migration and Quantum Security Verification Technologies) and NSF-BSF grant 2023627. \emph{(Corresponding author: Yongjune Kim.)}
}
}

\maketitle
\begin{abstract}
    
       Side-channel attacks (SCAs) pose a serious threat to system security by extracting secret keys through physical leakages such as power consumption, timing variations, and electromagnetic emissions. 
       Among existing countermeasures, artificial noise injection is recognized as one of the most effective techniques. 
       However, its high power consumption poses a major challenge for resource-constrained systems such as Internet of Things (IoT) devices, motivating the development of more efficient protection schemes.
       In this paper, we model SCAs as a communication channel and aim to suppress information leakage by minimizing the mutual information between the secret information and side-channel observations, subject to a power constraint on the artificial noise. 
       We first consider the Gaussian input case, where the mutual information becomes the channel capacity, which is one way to quantify the information leakage.
       We then extend the framework to arbitrary input distributions by identifying conditions under which the optimization remains convex and by leveraging the fundamental I–MMSE relationship to derive the optimal noise allocation. 
       Numerical results show that the proposed methods substantially reduce mutual information compared with conventional techniques, demonstrating their effectiveness for security-critical systems operating under tight power constraints.

    \end{abstract}  

    \begin{IEEEkeywords}
        Side-channel attacks, information leakage, mutual information, artificial noise injection.
    \end{IEEEkeywords}
    
\section{Introduction}\label{sec:introduction}
    
Side-channel attacks (SCAs) pose a serious threat to cryptographic security by targeting confidential information such as secret keys or messages through physical leakages, including power consumption, timing variations, and electromagnetic emissions~\cite{Brier2004correlation, Gandolfi2001electromagnetic, Jin2022optimal}. 
These attacks exploit variations in physical characteristics during secret-dependent operations and include power-analysis attacks~\cite{kocher1999differential}, timing attacks~\cite{kocher1996timing}, and electromagnetic attacks~\cite{al2016acoustic}. 
Among these, power-analysis attacks are especially effective, as they rely on monitoring a device's power consumption during cryptographic computations. 
By analyzing the resulting leakage traces, attackers aim to infer secret information~\cite{kocher1996timing, gattu2020power}.

To defend against SCAs, several countermeasures have been proposed, including artificial noise injection~\cite{shamir2000protecting}, masking~\cite{prouff2013masking}, and hiding techniques~\cite{veyrat2012shuffling}. 
Among these, artificial-noise injection aims to reduce the signal-to-noise ratio (SNR) of the observed leakage by adding external noise~\cite{shamir2000protecting, Jin2022optimal, das2018asni}. 
Although effective in principle, the approach is inefficient without careful allocation of artificial-noise power, as it typically requires noise power several times greater than the actual power consumption of encryption algorithms.

Such constraints are particularly pronounced in Internet of Things (IoT) devices, which are susceptible to SCAs due to their limited power and computational resources~\cite{Arsath2020param, Socha2022comprehensive}. 
While high-resource systems can implement artificial noise injection effectively, IoT devices often lack the capacity to support these high-overhead techniques. 
Consequently, they remain exposed to power-analysis attacks that exploit distinct power signatures during cryptographic operations. 

We propose an artificial-noise injection method that addresses the limitations of existing approaches by using mutual information~\cite{Cover2006} as a measure of side-channel information leakage.
Our approach formulates an optimization problem that minimizes the mutual information under a power constraint on the artificial noise, with particular focus on applicability to resource-constrained devices.
To formulate the optimization framework under a power constraint, we need a metric that accurately quantifies information leakage.
While maximal leakage~\cite{Issa2020maximal} is a well-known metric for quantifying information leakage, the mutual information is more appropriate in the context of power-analysis attacks, as it accommodates continuous random variables typically observed in side-channel measurements. 
In contrast, the maximal leakage diverges under these conditions~\cite[Example 10]{Issa2020maximal}.
    
We begin with the Gaussian input case, where the mutual information is equivalent to the channel capacity. 
This strategy can be interpreted as a \emph{dual water-filling} problem, which seeks to minimize the channel capacity to enhance security--contrasting with the classical water-filling approach that aims to maximize the total capacity of parallel Gaussian channels.
We then extend our optimization framework to arbitrary input distributions.
To derive the optimal solutions, we utilize the fundamental connection between mutual information and minimum mean squared error (MMSE), known as the I-MMSE relationship~\cite{Guo2005mutual}.
Compared to conventional uniform noise allocation, the proposed method significantly reduces the required artificial noise power while achieving the same level of leakage suppression.

   \section{System Model and Background}
    \subsection{Side-Channel Model}

    Suppose that the discrete uniform random variable $U$ denotes the secret information, such as a secret key or message, and the observed side-channel output signal, e.g., power consumption, is represented by the random vector $Y^m = \{Y_i\}_{i=1}^{m} = (Y_1, \ldots, Y_m ) \in \mathbb{R}^m$. 
    We introduce an intermediate random vector $X^m  \in \mathbb{R}^m$, defined as $X^m=g(U)$, where $g(\cdot)$ is a (possibly stochastic) function that models the side-channel of the secret information. 
    Let $Z^m$ and $N^m$ denote the physical and artificial noise vectors, respectively. 
    The side-channel observation model is then given by  
    \begin{equation}\label{eq:system_model}
        Y^m = g(U) + N^m + Z^m = X^m + N^m + Z^m,
    \end{equation}
    where the power of each intermediate component is denoted by $\mathbb{E}[X_i^2]=\mathsf{P}_i$. 
    Both noise vectors are modeled as Gaussian distributions; specifically, $Z_i \sim \mathcal{N}(0,\mathsf{Z}_i)$ and $N_i \sim \mathcal{N}(0,\mathsf{N}_i)$ for $i = 1, \ldots, m$. 
    The use of Gaussian noise is theoretically justified as it minimizes the mutual information between the input and output when the input is Gaussian \cite[Exercise 9.21]{Cover2006}.
    From a practical perspective, modeling noise as Gaussian is widely adopted, particularly for artificial noise in countermeasures against power side-channel attacks, as demonstrated in studies such as~\cite{Jin2022optimal, Gan2024classic, Guneysu2011generic}.
    
    \subsection{Information Leakage Metrics for SCA}
    
    In this subsection, we introduce representative information-leakage metrics and discuss related works on SCAs. 

    \subsubsection{Mutual Information}
    The mutual information measures the amount of information shared between two random variables. 
    In the context of SCAs, it quantifies the information about the secret variable $U$ revealed by the side-channel output $Y^m$. 
    It is defined as     
    \begin{equation} \label{eq:def_mutual_info}
         I(U; Y^m) = D_{\text{KL}}\left(P(u,y^m) || P(u)P(y^m)\right), 
    \end{equation}
    where $D_{\text{KL}}(P||Q)$ denotes the Kullback-Leibler divergence.
    Throughout this paper, $P(\cdot)$ denotes a general probability measure; $p(\cdot)$ and $f(\cdot)$ represent the probability mass function (pmf) of a discrete random variable and the probability density function (pdf) of a continuous random variable, respectively.
    The mutual information is widely adopted in both discrete and continuous settings for quantifying information leakage, due to its solid theoretical foundation and operational relevance in SCA evaluation~\cite{Gierlichs2008mutual,Jin2022optimal}.      
    Jin \emph{et al}.~\cite{Jin2022optimal} adopted the Gaussian channel capacity and formulated an optimization problem to control this capacity by injecting artificial noise that is \emph{uniformly allocated} across a subset of leakage points.      
    
    \subsubsection{Maximal Leakage}
    The maximal leakage quantifies the maximum gain in an adversary’s ability to guess the secret variable $U$ after observing the side-channel output $Y^m$, compared to making a blind guess~\cite{Issa2020maximal}. 
    Let $\hat{U}$ denote the estimate of $U$ inferred from $Y^m.$
    The maximal leakage is defined as 
    \begin{equation}\label{eq:def_maximal_leakage}
        L(X^m \rightarrow Y^m) = \sup_{U \rightarrow X^m \rightarrow  Y^m \rightarrow \hat{U}} \log \left( \frac{\Pr(U = \hat{U})}{\max_{u} P(u)} \right).
    \end{equation}
    While the maximal leakage provides a worst-case guarantee, it is not suitable for continuous variables, as it diverges in such cases~\cite{Issa2020maximal}.    
    In~\cite{Wu2020case}, the authors minimize the maximal leakage by injecting artificial delay in discrete time.

    \subsubsection{Sibson Mutual Information} 
    The Sibson mutual information is a generalization of mutual information, parameterized by a tunable order $\alpha > 0$~\cite{Sibson1969information,Verdu2015alpha}. 
    It is defined as 
    \begin{equation}\label{eq:sibson}
       I_\alpha (U;Y^m)=\inf_{P(y^m)} D_\alpha (P(u, y^m) || P(u)P(y^m)),
    \end{equation}
    where $D_\alpha(P||Q)$ denotes the R\'{e}nyi divergence of order $\alpha$. 
    As $\alpha \rightarrow 1$, $I_\alpha (U;Y^m)$ converges to the mutual information. 
    As $\alpha \rightarrow \infty$, it converges to the maximal leakage~\cite{Issa2020maximal}. 

    \subsection{I-MMSE Relation}
    
    To analyze the optimization problem involving the mutual information for arbitrary input distributions, we leverage one of the most fundamental relationships in information theory: the I-MMSE theorem~\cite{Guo2005mutual}. 
    As in~\cite{Guo2005mutual}, we rewrite $Y=X+N+Z$ in normalized form as
    \begin{equation}
        \widetilde{Y}=\sqrt{\rho}S+W,
    \end{equation}
    where $W$ represents the combined physical and artificial noise, and assume that $S$ and $W$ have unit power. 
    The signal-to-noise ratio (SNR) is given by  
    \begin{equation} \label{eq:snr}
    \rho = \frac{\mathsf{P}}{\mathsf{N}+\mathsf{Z}}. 
    \end{equation}    
    As in \cite{Lozano2006optimum}, regardless of their marginal distributions, any nonzero mean of input $S$ contributes to the power but not to the mutual information; hence, we assume zero-mean inputs without loss of generality.

    The mutual information between $X$ and $Y$ is expressed as a function of $\rho$~\cite{Guo2005mutual}:
    \begin{equation}
    I(X; Y) = I(S; \widetilde{Y})=I(\rho).
    \end{equation}
    \begin{lemma} [I-MMSE Relation~\cite{Guo2005mutual}]\label{thm:I-MMSE}
        For any input distribution, assuming a Gaussian $W$, the mutual information and MMSE satisfy
         \begin{equation}
         \frac{d}{d\rho}I(\rho)=\frac{1}{2}\mathsf{mmse}(\rho)=\frac{1}{2}\mathbb{E}\left[({S}-\mathbb{E}[{S}|{\widetilde{Y}};\rho])^2\right].
     \end{equation}
    \end{lemma}
    This relation enables the optimization of mutual information with respect to artificial noise allocation, even for arbitrary (non-Gaussian) inputs. 
    
    \section{Optimal Artificial Noise for Side-Channel Attack Resistance} \label{sec:optimization}
    
    An attacker attempts to infer the secret variable $U$ from the side-channel observation $Y^m$, and our objective is to increase the probability of inference error $P_e = \Pr(U \ne \hat{U})$. 
    According to Fano's inequality~\cite{Cover2006}, the probability of inference error satisfies 
    \begin{equation}
        1 + P_e \log{\lvert\mathcal{U}\rvert} \ge H(U|Y^m),
    \end{equation}
    where $\mathcal{U}$ denotes the alphabet of $U$.
    This inequality implies that a larger conditional entropy $H(U|Y^m)$ leads to a higher lower bound on $P_e$. 
    Since the mutual information is defined as $I(U;Y^m) = H(U) - H(U|Y^m)$, minimizing $I(U;Y^m)$ is equivalent to maximizing $H(U|Y^m)$, which in turn increases the lower bound on the probability of inference error. 
    Hence, we aim to minimize the mutual information $I(U;Y^m)$.
    
    We assume that the relationship between the secret information $U$ and the intermediate vector $X^m$ is immutable, and the system designer has control over how $Y^m$ is generated from $X^m$, as in~\cite{Wu2020case}.
    Hence, instead of directly minimizing the mutual information $I(U; Y^m)$, we aim to minimize its upper bound, which simplifies the optimization problem. 
    Since $U \rightarrow X^m \rightarrow Y^m$ forms a Markov chain, the following first inequality holds by the data-processing inequality, and the second follows from the chain rule: 
    \begin{equation}\label{eq:data_processing_inequality}
         I(U;Y^m)\le I(X^m;Y^m) \le \sum_{i=1}^{m} I(X_i;Y_i). 
    \end{equation}
    Moreover, this data-processing inequality also holds for the Sibson mutual information~\cite{Verdu2015alpha}, making it applicable in more general settings involving $\alpha$-parametrized leakage measures.

    We aim to minimize information leakage by injecting artificial noise under a fixed artificial-noise power budget. 
    As information-leakage metrics, we adopt the mutual information and the Sibson mutual information. 
    However, the maximal leakage is not suitable in our setting as $Y=X+N+Z$.
    Note that $L(X^m \rightarrow Y^m) \rightarrow \infty$ for an additive model with continuous variables~\cite[Example 10]{Issa2020maximal}.
        
    We can then formulate the following optimization problem: 
    \begin{align} \label{eq:main_opt}
    &\underset{{\{\mathsf{N}_i\}_{i=1}^m}}{\text{minimize}}~\sum_{i=1}^{m} I(X_i;Y_i) \\
        &\text{subject to} ~\sum_{i=1}^m \mathsf{N}_i\le \mathsf{N}_0, \quad \mathsf{N}_i\ge0, \nonumber 
    \end{align}
    where  $\mathsf{N}_0$ denotes the total artificial-noise power constraint. 
    Note that from~\eqref{eq:system_model} each mutual information term $I(X_i;Y_i)$ is a function of $\mathsf{N}_i$ for $i = 1, \ldots, m$.
    
    \subsection{Gaussian Input Distribution}\label{sec:gaussian_average}
    
    Similarly to~\cite{Jin2022optimal}, we adopt a sub-channel perspective, where each leakage point in the side-channel is modeled as a distinct sub-channel.
    The amount of leakage is quantified by the sub-channel powers, $\mathsf{P}_i$ for $i=1,\ldots,m$, and is assumed to be known. 
    In addition, we assume that the variances $\mathsf{Z}_i$ of the physical noise are known. 
    We focus on the case where each input $X_i$ follows a Gaussian distribution and aim to minimize the mutual information.  
    For an additive Gaussian-noise channel, this mutual information equals the channel capacity. Hence, the optimization problem becomes:
    \begin{align}\label{eq:capacity_opt}
        &\underset{{\{\mathsf{N}_i\}_{i=1}^m}}{\text{minimize}}~ \sum_{i=1}^{m} \frac{1}{2}\log\left(1+\frac{\mathsf{P}_i}{\mathsf{N}_i+\mathsf{Z}_i}\right) \\
        &\text{subject to}~ \sum_{i=1}^m \mathsf{N}_i\le \mathsf{N}_0, \quad \mathsf{N}_i\ge0. \nonumber
    \end{align}

    \begin{remark}\label{thm:capacity_convexity} 
    The optimization problem in~\eqref{eq:capacity_opt} is \emph{convex} with respect to $\{\mathsf{N}_i\}_{i=1}^m$ since the second derivative of the objective function is non-negative.        
    \end{remark}

    The optimal solution can be obtained using standard convex optimization solvers.
    Additionally, we provide an analytical characterization of the optimal solution by applying the KKT conditions.
    
    \begin{theorem}\label{thm:gaussian}
    The optimal noise allocation $\{\mathsf{N}_i^*\}_{i=1}^m$ of \eqref{eq:capacity_opt} is given by
    \begin{equation}\label{eq:gaussian_optimal_noise}    
        \mathsf{N}_i^*=\begin{cases} 
            0, & \nu \ge \frac{1}{\mathsf{Z}_i} - \frac{1}{\mathsf{Z}_i + \mathsf{P}_i}, \\
            \frac{-(2\mathsf{Z}_i+\mathsf{P}_i)+\sqrt{\mathsf{P}_i^2+\frac{4 \mathsf{P}_i}{\nu}}}{2}, & \text{otherwise},            
        \end{cases}
    \end{equation}
    where $\nu$ is the dual variable chosen to satisfy the condition $\sum_{i=1}^m\mathsf{N}_i=\mathsf{N}_0$. 
    Also, for $\mathsf{N}_i^* > 0$, the optimal $\mathsf{N}_i^*$ satisfies the following condition: 
    \begin{equation}\label{eq:gaussian_nu}
       \frac{1}{\mathsf{N}_i^* + \mathsf{Z}_i} - \frac{1}{\mathsf{N}_i^* + \mathsf{Z}_i + \mathsf{P}_i}=\nu.
    \end{equation}    
    \begin{IEEEproof}
    We consider only the case where $\mathsf{P}_i > 0$, as the mutual information becomes zero when $\mathsf{P}_i = 0$, indicating the absence of information leakage. 
    In such cases, allocating artificial noise power is unnecessary, and the optimal noise allocation is naturally zero. 
    Therefore, we focus on components with $\mathsf{P}_i > 0$. 
    We define the Lagrangian $\mathcal{L}_1$ associated with~\eqref{eq:capacity_opt} as follows: 
    \begin{align}\label{eq:capacity_lagrangian}
        \mathcal{L}_1&=\sum_{i=1}^m{\log\left(1 + \frac{ \mathsf{P}_i }{\mathsf{N}_i+\mathsf{Z}_i }\right)}\nonumber\\&-\sum_{i=1}^m\lambda_i\mathsf{N}_i+\nu\left(\sum_{i=1}^m \mathsf{N}_i-\mathsf{N}_0\right),  
    \end{align}
    where $\lambda_i$ and $\nu$ are the dual variables.  
    We have the following KKT conditions: 
    \begin{align}
    &\lambda_i\ge0,\quad\nu\ge0,\quad \frac{\partial \mathcal{L}_1}{\partial\mathsf{N}_i} = 0, \label{eq:gaussian_kkt0} \\
    &\lambda_i\cdot\mathsf{N}_i=0,\quad\nu\left(\sum_{i=1}^m\mathsf{N}_i-\mathsf{N}_0\right)=0.\label{eq:gaussian_kkt1}
    \end{align}
    From the KKT conditions, we obtain
    \begin{equation}\label{eq:gaussian_lambda}
        \lambda_i=\frac{1}{\mathsf{N}_i+\mathsf{Z}_i+\mathsf{P}_i}-\frac{1}{\mathsf{N}_i+\mathsf{Z}_i} +\nu\ge0.
    \end{equation}
    If $\nu=0$, then $\lambda_i < 0$ for $\mathsf{P}_i > 0$, which violates the KKT conditions $\lambda_i\ge0$. 
    Hence, we should have $\nu > 0$, and by \eqref{eq:gaussian_kkt1}, $\sum_{i=1}^m\mathsf{N}_i=\mathsf{N}_0$. 
    The condition $\lambda_i\cdot\mathsf{N}_i=0$ of \eqref{eq:gaussian_kkt1} leads to
    \begin{equation} \label{eq:slackness_gaussian}
        \mathsf{N}_i\left(\frac{1}{\mathsf{N}_i+\mathsf{Z}_i+\mathsf{P}_i}-\frac{1}{\mathsf{N}_i+\mathsf{Z}_i}+\nu\right)=0.
    \end{equation}

    We now consider three cases based on the value of $\nu$ relative to the threshold $\frac{1}{\mathsf{Z}_i} - \frac{1}{\mathsf{Z}_i + \mathsf{P}_i}$: 
    \begin{itemize}
        \item If $\nu > \frac{1}{\mathsf{Z}_i} - \frac{1}{\mathsf{Z}_i + \mathsf{P}_i}$, then we have $\lambda_i > 0$ by \eqref{eq:gaussian_lambda}, and thus $\mathsf{N}_i = 0$ by \eqref{eq:gaussian_kkt1}. 
        \item If $\nu = \frac{1}{\mathsf{Z}_i} - \frac{1}{\mathsf{Z}_i + \mathsf{P}_i}$, then \eqref{eq:slackness_gaussian} holds for $\mathsf{N}_i = 0$.
        \item If $\nu < \frac{1}{\mathsf{Z}_i} - \frac{1}{\mathsf{Z}_i + \mathsf{P}_i}$, then setting $\mathsf{N}_i = 0$ leads to $\lambda_i < 0$ in \eqref{eq:gaussian_lambda}, which violates \eqref{eq:gaussian_kkt0}. 
        Hence, we should have $\mathsf{N}_i > 0 $ and $\lambda_i = 0$, which leads to \eqref{eq:gaussian_nu}. 
        Solving \eqref{eq:gaussian_nu} yields the optimal solution in \eqref{eq:gaussian_optimal_noise}.
    \end{itemize}
    By combining these three cases, we obtain the optimal solution given in \eqref{eq:gaussian_optimal_noise}.
    \end{IEEEproof}
    \end{theorem}
    
    Our optimization problem in \eqref{eq:capacity_opt} can be viewed as a \emph{dual} to the classical water-filling problem~\cite{Cover2006}, which maximizes the total capacity of the parallel Gaussian channels by allocating transmit power $\{\mathsf{P}_i\}_{i=1}^m$ across channels. 
    In the water-filling problem, artificial noise is not considered; the optimization variables are $\{\mathsf{P}_i\}_{i=1}^m$ for a given set of noise powers $\{\mathsf{Z}_i\}_{i=1}^m$. 
    For the selected channels, the optimal power allocation satisfies the condition $\mathsf{P}_i^* +\mathsf{Z}_i = \nu$, where $\nu$ represents the water-level. 
    In contrast, our objective is to minimize the total capacity of parallel Gaussian channels by allocating artificial noise $\{\mathsf{N}_i\}_{i=1}^m$. 
    For the selected points (i.e., those with non-zero artificial noise allocation), the optimal noise allocation $\mathsf{N}_i^*$ satisfies the condition in \eqref{eq:gaussian_nu}, which serves as the dual counterpart to the optimal $\mathsf{P}_i^*$ in the water-filling solution. 

    The optimization problem can be extended to the setting of Sibson mutual information. 
    According to~\cite{Verdu2015alpha}, for a Gaussian channel with a Gaussian input, the Sibson mutual information for $\alpha > 0$ is given by 
    \begin{equation} \label{eq:alpha_capacity}
    I_{\alpha}(X_i;Y_i) = \frac{1}{2}\log\left(1 + \alpha \cdot\frac{\mathsf{P}_i}{\mathsf{N}_i+\mathsf{Z}_i}\right).
    \end{equation}
    Substituting \eqref{eq:alpha_capacity} into \eqref{eq:main_opt} yields a variant of \eqref{eq:capacity_opt} with the SNR scaled by $\alpha$, which remains convex as in Remark~\ref{thm:capacity_convexity}.
    The optimal solution can be derived by applying the KKT conditions.
        
    \begin{corollary}
    The optimal noise allocation $\{\mathsf{N}_i^*\}_{i=1}^m$ for the optimization problem with the Sibson mutual information is given by:
    \begin{equation}
        \mathsf{N}_i^* = \begin{cases} 
            0, &\nu \ge \frac{1}{\mathsf{Z}_i} - \frac{1}{\mathsf{Z}_i + \alpha\mathsf{P}_i}, \\
            \frac{-(2\mathsf{Z}_i+\alpha \mathsf{P}_i) + \sqrt{(\alpha \mathsf{P}_i)^2 + \frac{4 \alpha \mathsf{P}_i}{\nu}}}{2}, &\text{otherwise},
        \end{cases}
    \end{equation}
    where $\nu$ is the dual variable. Also, for $\mathsf{N}_i^*>0$, the noise allocation $\{\mathsf{N}_i^*\}_{i=1}^m$ satisfies the following condition:
    \begin{equation}
        \frac{1}{\mathsf{N}_i^*+\mathsf{Z}_i} - \frac{1}{\mathsf{N}_i^*+\mathsf{Z}_i+\alpha \mathsf{P}_i} = \nu.
    \end{equation}
    \end{corollary}

    Our framework is flexible and adaptable to generalized information leakage measures, where $\alpha$ serves as a tunable parameter that modulates the sensitivity to signal power. 

    \begin{remark}
    As $\alpha \to \infty$, the Sibson mutual information converges to the maximal leakage~\cite{Issa2020maximal}. 
    However, in the case of Gaussian inputs and additive Gaussian noise, the maximal leakage becomes unbounded, since \eqref{eq:alpha_capacity} diverges as $\alpha \to \infty$. 
    Therefore, the maximal leakage is not suitable for Gaussian settings.
    \end{remark}
    \subsection{Arbitrary Input Distribution}\label{sec:arbitrary_average}

    In this subsection, we investigate the optimization problem in \eqref{eq:main_opt} under the assumption that the input signals follow an arbitrary distribution.
    Although our objective is to derive an optimal solution for arbitrary input distributions, the optimization problem in general is not convex in this setting.
    To facilitate tractable optimization, we first establish sufficient conditions under which the problem becomes convex.
    
    \begin{theorem}\label{thm:convex_condition}
        The mutual information $I(S;\widetilde{Y})$ (equiv. $I(X; Y)$) is convex in $\mathsf{N}$ if and only if one of the following (equivalent) conditions is satisfied for the variable $\rho = \frac{\mathsf{P}}{\mathsf{N}+\mathsf{Z}} > 0$. 
        \begin{align}
            &(\text{C1})~ \mathsf{mmse}\left(\rho\right)+\frac{\partial}{\partial\rho}\left(\rho\cdot\mathsf{mmse}(\rho)\right)\ge0,\label{eq:C1}\\
            &(\text{C2})~ 
            \frac{d}{d\rho}\left(\rho\cdot \mathcal{J}(\widetilde{Y})\right)\le 1,\label{eq:C2}\\
            &(\text{C3})~ 
            \mathbb{E}\left[\left(\frac{d^2}{d\widetilde{Y}^2}\log f(\widetilde{Y})\right)^2\right]\le1,\label{eq:C3}
        \end{align}
        where $\mathcal{J}(\widetilde{Y})=\mathbb{E}\left[\left(\frac{d}{d\widetilde{Y}}\log f(\widetilde{Y})\right)^2\right]$ is the Fisher information.
 
    \end{theorem}
    \begin{IEEEproof}
        To prove convexity, we derive the second derivative of $I(\rho)$. 
       
    \begin{align}
    &\frac{\partial^2}{\partial \mathsf{N}^2}I\left(\rho\right)\nonumber\\&=\frac{\partial}{\partial\mathsf{N}}\frac{d\rho}{d\mathsf{N}}\left(\frac{1}{2}\mathsf{mmse}(\rho)\right)\label{eq:second_derivative_MI0}
    \\&=-\frac{\partial}{\partial\mathsf{N}}\left(\frac{\mathsf{P}}{2\left(\mathsf{N}+\mathsf{Z}\right)^2}\mathsf{mmse}(\rho)\right)\label{eq:second_derivative_MI1}\\
    &=\frac{\mathsf{P}}{2\left(\mathsf{N}+\mathsf{Z}\right)^3}\left(\mathsf{mmse}(\rho) + \frac{d}{d\rho} \left( \rho \cdot \mathsf{mmse}(\rho) \right)\right),\label{eq:second_derivative_MI2}
    \end{align}
    where \eqref{eq:second_derivative_MI0} follows from the I-MMSE relation and the chain rule.

   \emph{Proof of (C1)}:
   From \eqref{eq:second_derivative_MI2} we have a factor $\frac{\mathsf{P}}{\left(\mathsf{N}+\mathsf{Z}\right)^3}$ which is positive, (C1) holds.

    \emph{Proof of (C2)}: 
    As shown in \cite{Guo2005mutual}, Fisher information and MMSE satisfy
    \begin{equation}\label{eq:fisher}
         \mathcal{J}(\widetilde{Y}) =\mathbb{E}\left[\left(\left.\frac{d}{dy}\log f_{\widetilde{Y}}(y)\right|_{y=\widetilde{Y}}\right)^2\right]= 1 - \rho\cdot \mathsf{mmse}(\rho).
    \end{equation}
    Combining \eqref{eq:second_derivative_MI2} and \eqref{eq:fisher}, 
   \begin{equation}\label{eq:con2_fisher}
    \frac{\partial^2}{\partial \mathsf{N}^2}I\left(\rho\right)= \frac{\mathsf{1}}{2\left(\mathsf{N}+\mathsf{Z}\right)^2} \left( 1 - \mathcal{J}(\widetilde{Y}) - \rho \frac{d}{d\rho} \mathcal{J}(\widetilde{Y}) \right).
    \end{equation}
    Since $\frac{\mathsf{1}}{\left(\mathsf{N}+\mathsf{Z}\right)^2}$ is always positive, (C2) holds.
    
    \emph{Proof of (C3)}: According to~\cite{Guo2011mmse}, the conditional variance is defined as 
    \begin{equation}
        M = \mathrm{Var}(S | \widetilde{Y}) = \mathbb{E} \left[ \left( S - \mathbb{E} \{ S | \widetilde{Y} \} \right)^2 \Big|\widetilde{Y} \right].
    \end{equation}
    In addition, \cite{Guo2011mmse} provides some properties of MMSE with conditional variance. 
    \begin{align}
        \mathsf{mmse}(\rho) &= \mathbb{E}[M]\label{eq:mmse_variance1},\\\frac{d}{d\rho} \mathsf{mmse}(\rho) &= -\mathbb{E}[M^2].\label{eq:mmse_variance2}
    \end{align}

    Also, $M$ can be represented as follows using Tweedie's formula~\cite{Efron2011tweedie}: 
    \begin{equation}\label{eq:Tweedies}
         M= \frac{1}{\rho} \left( 1 + \frac{d^2}{dy^2} \log f_{\widetilde{Y}}(y) \right).
    \end{equation}
    Combining \eqref{eq:second_derivative_MI2}, \eqref{eq:mmse_variance1}, \eqref{eq:mmse_variance2}, and \eqref{eq:Tweedies}, we have,
    \begin{align}
        \frac{\partial^2}{\partial \mathsf{N}^2}I\left(\rho\right)= \frac{\mathsf{P}}{2\left(\mathsf{N}+\mathsf{Z}\right)^2} \mathbb{E} \left[ 1 - \left( \left.\frac{d^2}{dy^2} \log f_{\widetilde{Y}}(y) \right|_{y=\widetilde{Y}}\right)^2 \right].
    \end{align}
    Since $\frac{\mathsf{P}}{\left(\mathsf{N}+\mathsf{Z}\right)^2}$ is always positive, (C3) holds.
    \end{IEEEproof}

    \begin{example}[Gaussian Input] \label{example:Gaussian}
    Consider the case where both the input and the noise are Gaussian. In this case, $\widetilde{Y} \sim \mathcal{N}(0, 1 + \rho)$, which leads to $\mathsf{mmse}(\rho)=\frac{1}{1+\rho}$.
    Substituting this into (C1), we verify that the convexity conditions are satisfied.
    \end{example}
    
    \begin{example}[Binary Input]\label{example:Binary}
    The MMSE for binary input $S\in \{\pm 1 \}$ with equal probability is given in \cite{Lozano2006optimum} as follows:
    \begin{equation}
        \mathsf{mmse}(\rho) = 1 - \int_{-\infty}^{\infty} 
    \frac{e^{-{y}^{2}/2}}{\sqrt{2\pi}} \tanh\!\bigl(\rho - \sqrt{\rho}\, {y}\bigr) \, dy .
    \end{equation}
    We observe that the binary input violates (C1); in particular, \eqref{eq:C1} is negative when $\rho >3.35$.
    Thus, the optimization problem \eqref{eq:main_opt} is non-convex. 
    \end{example}

    \begin{example}[Exponential Input]\label{example:exponential}
    For an exponential input distribution, the output follows an exponentially modified Gaussian distribution \cite{Grushka1972characterization}, which satisfies (C3). 
    Thus, the optimization problem \eqref{eq:main_opt} is convex. 
    \begin{IEEEproof}
     According to the well-known \emph{exponentially modified Gaussian distribution}~\cite{Grushka1972characterization}, the distribution of $Y$ has the following pdf:
 \begin{align}
    f_{Y}(y; \beta,\sigma)=\beta\exp\left(\frac{\beta^2\sigma^2}{2}-\beta y\right) Q\left(\beta\sigma-\frac{y}{\sigma}\right),
    \end{align}
where $Q(x) = \int_{x}^{\infty}{\frac{1}{\sqrt{2\pi}} \exp \left( - \frac{t^2}{2} \right) dt}$. 

    Define $\widetilde{Y}\triangleq Y/\sigma$; from elementary probability we have $f_{\widetilde{Y}}(y; \beta,\sigma)=\sigma f_{Y}(\sigma y; \beta,\sigma)$. Thus
 \begin{align}
        f_{\widetilde{Y}}(y; \beta,\sigma)=\beta\sigma\exp\left(\frac{\beta^2\sigma^2}{2}-\beta\sigma y\right) Q\left(\beta\sigma-y\right). \label{eq:pdf_exp_mod}
    \end{align}
We now define the SNR $\rho\triangleq 2/(\beta\sigma)^2$ where $\mathbb{E}[X^2]=\frac{2}{\beta^2}$ for $X\sim Exp(\beta)$, and substitute it into~\eqref{eq:pdf_exp_mod}:
\begin{align}
        &f_{\widetilde{Y}}(y; \rho)=\sqrt{\frac{2}{\rho}}\exp\left(\frac{1}{\rho}-\sqrt{\frac{2}{\rho}}y\right) Q\left(\sqrt{\frac{2}{\rho}} -y\right). \label{eq:pdf_exp_mod_sub_old}
    \end{align}

Note that~\eqref{eq:pdf_exp_mod_sub_old} is only a function of the SNR $\rho$. 
 We aim to prove that exponential input satisfies (C3).
    For simplicity, let $v \triangleq -{y} + \sqrt{\frac{2}{\rho}}$ and define $h(v)=\frac{d^2}{d{y}^2}\log f_{\widetilde{Y}}({y};\rho)$. 
    Then, $h(v)$ can be represented as follows: 
    \begin{align}
        h(v)
        &=-\frac{Q'(v)}{Q(v)}\left(v+\frac{Q'(v)}{Q(v)}\right)\label{eq:exponential_h_3}.
    \end{align}
    
    To prove $\mathbb{E}[h(v)^2] \le 1$, it suffices to show that $-1\le h(v)\le0$ for all $v$.
    We first verify that $h(v)\le0$ for all $v$. 
    For $v \le 0$, it is straightforward since ${Q(v)}>0, {Q'(v)}<0$. 
    For $v>0$, the $Q$-function can be rewritten as follows in \cite{Jacobs1965principles}:
    \begin{align}
        &\sqrt{2\pi}Q(v)\nonumber\\&=\int_{v}^\infty\frac{1}{t}\cdot t\cdot\exp\left(-\frac{t^2}{2}\right)dt
        \\&=\exp\left(-\frac{{v}^2}{2}\right)\frac{1}{{v}}-\int_{v}^\infty\frac{1}{t^2}\cdot \exp\left(-\frac{t^2}{2}\right)dt.  \label{eq:Q-function_taylor_expansion_1}
    \end{align}
    Since the last term of \eqref{eq:Q-function_taylor_expansion_1} is always positive, the $Q$-function satisfies \eqref{eq:q_func_inequality}. Substituting \eqref{eq:q_func_inequality} into \eqref{eq:exponential_h_3} yields $h(v)\le 0$. 
    \begin{align} \label{eq:q_func_inequality}
    Q(v)&<\frac{1}{\sqrt{2\pi}}\exp\left(-\frac{{v}^2}{2}\right)\frac{1}{{v}}=-\frac{Q'(v)}{v}.
    \end{align}
    
    Next, we show that $h(v)\ge-1$. 
    Using $Q'(v)=-\phi(v)$, where $\phi(v)$ is the standard normal probability density function, we obtain the following:
    \begin{align}
    h(v) + 1=\frac{ Q(v)^2 + v\,\phi(v)\,Q(v) - \phi(v)^2 }{ Q(v)^2 }\label{eq:Q-fucnntion_h+1}.
    \end{align}
    Let $k(v)$ denote the numerator of \eqref{eq:Q-fucnntion_h+1}. 
    The sign of $h(v)+1$ is the same as that of $k(v)$. 
    To prove $k(v)\ge0$ for all $v$, we argue by contradiction: if $k(v)$ were negative at some points, it would contradict the required condition. 
    To examine the behavior of \(k(v)\), we evaluate its limiting values, which follow directly from \eqref{eq:Q-fucnntion_h+1}.
     \begin{align}
        \lim _{v\rightarrow-\infty} k(v)=1, \quad \lim _{v\rightarrow\infty}k(v)=0.\label{eq:limit_k(v_i)}
    \end{align} 
    
    If $k(v)$ attains a negative value at some point, then there should exist a point $v^*\in(-\infty,\infty)$ such that $k'(v^*)=0$, as guaranteed by the intermediate value theorem and Rolle's theorem~\cite{Apostol1991calculus}.
    \begin{equation}
        k'(v^*) = \phi(v^*) \left\{ v^*\,\phi(v^*) - (1+{v^*}^2)\,Q(v^*) \right\}=0.
    \end{equation}
    Since $\phi(\cdot)$ is always positive, $v^*$ satisfies
    \begin{equation}\label{eq:Q-function_derivative_k}
        v^*\,\phi(v^*) = (1+{v^*}^2)\,Q(v^*),
    \end{equation}
    which results in $v>0$.
    For $v>0$, the $Q$-function should satisfy the Mills ratio \cite{Gordon1941values}: 
    \begin{equation}\label{eq:mills_ratio}
        Q(v)>\left(\frac{v}{1+{v}^2} \right)\phi(v).
    \end{equation}
    Since \eqref{eq:Q-function_derivative_k} would contradict \eqref{eq:mills_ratio}, $k(v)$ is nonnegative for all $v$, which implies $h(v) \ge -1$. 
    Combining this with the earlier result $h(v) \le 0$, we conclude that $h(v) \in [-1,0]$. 
    Thus, the exponential input distribution satisfies (C3) of Theorem~\ref{thm:convex_condition}.   
    \end{IEEEproof}    
    \end{example}
    
    If the optimization problem \eqref{eq:main_opt} is convex, we can obtain the global optimal solution using the I-MMSE relation and the KKT conditions.
    \begin{theorem}\label{thm:arbitrary_optimal} 
    For an arbitrary input distribution that satisfies conditions in Theorem~\ref{thm:convex_condition}, the optimal noise allocation $\{\mathsf{N}_i\}_{i=1}^m$ is given by
    \begin{align}
        \mathsf{N}_i & =0,\quad\text{if}\quad \nu \ge \delta_i,\label{eq:arbitrary_optimal_noise1}\\
        \frac{\mathsf{P}_i}{(\mathsf{N}_i+\mathsf{Z}_i)^2}\cdot \mathsf{mmse}\left(\frac{\mathsf{P}_i}{\mathsf{N}_i+\mathsf{Z}_i}\right)&=\nu,\quad\text{otherwise} ,\label{eq:arbitrary_optimal_noise2}
    \end{align}
    where $\nu$ is a dual variable of the corresponding KKT conditions and $\delta_i=\frac{\mathsf{P}_i}{\mathsf{Z}_i^2}\cdot\mathsf{mmse}\left(\frac{\mathsf{P}_i}{\mathsf{Z}_i}\right)$.
    \end{theorem}
    \begin{IEEEproof}
        We consider only the case where $\mathsf{P}_i > 0$, for the same reason as in the proof of Theorem~\ref{thm:gaussian}.
 We define the Lagrangian $\mathcal{L}_2$ associated with~\eqref{eq:main_opt} as follows:
\begin{align}
    \mathcal{L}_2
    &=\sum_{i=1}^mI\left(\frac{\mathsf{P}_i}{\mathsf{N}_i+\mathsf{Z}_i}\right)-\sum_{i=1}^m\lambda_i\mathsf{N}_i \nonumber \\
    &+\nu\left( \sum_{i=1}^m \mathsf{N}_i - \mathsf{N}_0\right),
\end{align}
where $\lambda_i,\nu$ are dual variables. We have the following KKT conditions:
\begin{align}
&\lambda_i\ge0,\quad\nu\ge0, \quad\frac{\partial \mathcal{L}_2}{\partial \mathsf{N}_i}=0, \label{eq:arbitrary_kkt1}\\
&\lambda_i\cdot\mathsf{N}_i=0,\quad\nu\left(\sum_{i=1}^m\mathsf{N}_i-\mathsf{N}_0\right)=0. \label{eq:arbitrary_kkt2} 
\end{align}
From \eqref{eq:arbitrary_kkt1}, we obtain 
\begin{equation}\label{eq:arbitrary_lambda}
    \lambda_i=-\frac{\mathsf{P}_i}{(\mathsf{N}_i+\mathsf{Z}_i)^2}\cdot \mathsf{mmse}\left(\frac{\mathsf{P}_i}{\mathsf{N}_i+\mathsf{Z}_i}\right)+\nu.
\end{equation}
If $\nu=0$, then $\lambda_i<0$ for $\mathsf{P}_i>0$, which violates the KKT conditions $\lambda_i\ge0$. 
Hence, we should have $\nu>0$, and by \eqref{eq:arbitrary_kkt2}, $\sum_{i=1}^m\mathsf{N}_i=\mathsf{N}_0$.
The condition $\lambda_i\cdot\mathsf{N}_i=0$ of \eqref{eq:arbitrary_kkt2} leads to 
\begin{equation}\label{eq:slackness_arbitrary}
    \mathsf{N}_i\left(-\frac{\mathsf{P}_i}{(\mathsf{N}_i+\mathsf{Z}_i)^2}\cdot \mathsf{mmse}\left(\frac{\mathsf{P}_i}{\mathsf{N}_i+\mathsf{Z}_i}\right)+\nu\right)=0.
\end{equation}

We now consider three cases based on the value of $\nu$ relative to the threshold $\frac{\mathsf{P}_i}{\mathsf{Z}_i^2}\cdot\mathsf{mmse}\left(\frac{\mathsf{P}_i}{\mathsf{Z}_i}\right)$: 
    \begin{itemize}
        \item If $\nu > \frac{\mathsf{P}_i}{\mathsf{Z}_i^2}\cdot\mathsf{mmse}\left(\frac{\mathsf{P}_i}{\mathsf{Z}_i}\right)$, then we have $\lambda_i > 0$ by \eqref{eq:arbitrary_lambda}, since $\frac{\mathsf{P}_i}{(\mathsf{N}_i+\mathsf{Z}_i)^2}\cdot \mathsf{mmse}\left(\frac{\mathsf{P}_i}{\mathsf{N}_i+\mathsf{Z}_i}\right)$ is a decreasing function for $\mathsf{N}_i$ by Remark \ref{remark:decreasing_nu}. 
        Therefore, $\mathsf{N}_i = 0$ by \eqref{eq:arbitrary_kkt2}. 
        \item If $\nu = \frac{\mathsf{P}_i}{\mathsf{Z}_i^2}\cdot\mathsf{mmse}\left(\frac{\mathsf{P}_i}{\mathsf{Z}_i}\right)$, then \eqref{eq:slackness_arbitrary} holds for $\mathsf{N}_i = 0$.
        \item If $\nu < \frac{\mathsf{P}_i}{\mathsf{Z}_i^2}\cdot\mathsf{mmse}\left(\frac{\mathsf{P}_i}{\mathsf{Z}_i}\right)$, then setting $\mathsf{N}_i = 0$ leads to $\lambda_i < 0$ in \eqref{eq:arbitrary_lambda}, which violates \eqref{eq:arbitrary_kkt1}. 
        Hence, we should have $\mathsf{N}_i>0$ and $\lambda_i=0$, which leads to 
\begin{equation}
    \frac{\mathsf{P}_i}{(\mathsf{N}_i+\mathsf{Z}_i)^2}\cdot \mathsf{mmse}\left(\frac{\mathsf{P}_i}{\mathsf{N}_i+\mathsf{Z}_i}\right)=\nu.
\end{equation}
  \end{itemize}
By combining these three cases, we obtain the optimal solution given in Theorem~\ref{thm:arbitrary_optimal}.
    \end{IEEEproof}
    
    \begin{remark}\label{remark:decreasing_nu}
        If $I(\rho_i)$ is convex in $\mathsf{N}_i$, then $\nu = \frac{\mathsf{P}_i}{(\mathsf{N}_i+\mathsf{Z}_i)^2}\cdot \mathsf{mmse}\left(\frac{\mathsf{P}_i}{\mathsf{N}_i+\mathsf{Z}_i}\right)$ is nonincreasing in $\mathsf{N}_i$ because this quantity equals $-\frac{\partial}{\partial\mathsf{N}_i}I(\rho_i)$.
    \end{remark}
    
    The optimal artificial noise allocation can be obtained through the following three-step procedure:

    \begin{enumerate}
        \item Compute the MMSE function corresponding to the given input distribution.
        \item Solve for the dual variable $\nu$ using ~\eqref{eq:arbitrary_optimal_noise1} and \eqref{eq:arbitrary_optimal_noise2} under the constraint $\sum_{i=1}^m \mathsf{N}_i= \mathsf{N}_0$. 
        This step can be carried out efficiently using a bisection method by leveraging Remark~\ref{remark:decreasing_nu}.
        \item Determine the optimal noise allocation $\{\mathsf{N}_i\}_{i=1}^m$ based on the computed value of $\nu$.
    \end{enumerate}

    \begin{example}[Gaussian Input]
    We can obtain the optimal solution for a Gaussian input distribution using Theorem~\ref{thm:arbitrary_optimal}. The MMSE of a Gaussian input is $        \mathsf{mmse}\left(\frac{\mathsf{P}_i}{\mathsf{N}_i+\mathsf{Z}_i}\right)=\frac{\mathsf{N}_i+\mathsf{Z}_i}{\mathsf{N}_i+\mathsf{Z}_i+\mathsf{P}_i}$.    
    Using Theorem~\ref{thm:arbitrary_optimal}, the optimal solution can be derived as follows: 
    \begin{align}
            \mathsf{N}_i^*&=0,\ \text{if}\ \ \nu \ge \frac{1}{\mathsf{Z}_i} - \frac{1}{\mathsf{Z}_i + \mathsf{P}_i},\\
            \frac{1}{\mathsf{N}_i^* + \mathsf{Z}_i} - \frac{1}{\mathsf{N}_i^* + \mathsf{Z}_i + \mathsf{P}_i}&=\nu,\ \text{otherwise}.\label{eq:arbitrary_gaussian}
        \end{align}
    Note that \eqref{eq:arbitrary_gaussian} is identical to \eqref{eq:gaussian_nu} of Theorem~\ref{thm:gaussian}. 
    \end{example}

    For a broad class of input distributions, the MMSE function has been extensively studied in the context of Gaussian channels~\cite{Lozano2006optimum}, thereby enabling the computation of the corresponding optimal solution.   
    
    \subsection{Low-SNR Regime}
    
    An analysis of the optimization problem in the low-SNR regime reveals useful properties.
    Specifically, as the SNR approaches zero, the mutual information of the Gaussian channel is convex in the artificial noise for any input distribution.
    Following the analysis in~\cite{Guo2011mmse}, the second-order Taylor series expansion of the MMSE for a Gaussian channel around $\rho=0^+$ is given by  
    \begin{equation}\label{eq:low_snr_mmse}
        \mathsf{mmse}(\rho)=1-\rho+\left(2-\mathbb{E}[S^3]^2\right)\frac{\rho^2}{2}+\mathcal{O}(\rho^3).
    \end{equation}
    This expansion satisfies (C1) of Theorem~\ref{thm:convex_condition}, indicating that the objective function is always convex in the low-SNR regime. 
    Consequently, the optimization problem remains convex regardless of the input distribution, ensuring that optimal solutions can be obtained for both Gaussian and non-Gaussian inputs.
    Note that the SCA scenario typically corresponds to the low-SNR regime. 
    Even for the binary input signal in Example~\ref{example:Binary}, the problem is convex when $\rho < 3.35 \approx \SI{5.25}{dB}$, which lies in the low-SNR regime. 
    
    \section{Numerical Results}\label{sec:experiments}
 \begin{figure}[t]
        \centering
        \subfloat[]{\includegraphics[width=0.33\textwidth]{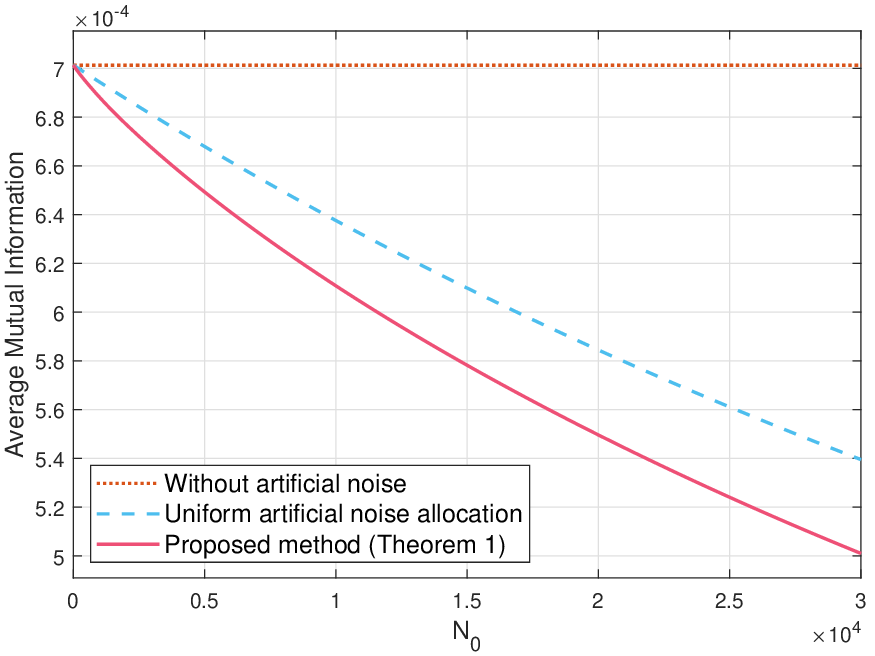}}
        \label{Fig: Gaussian_low_snr}
        \hfill
        \subfloat[]{\includegraphics[width=0.33\textwidth]{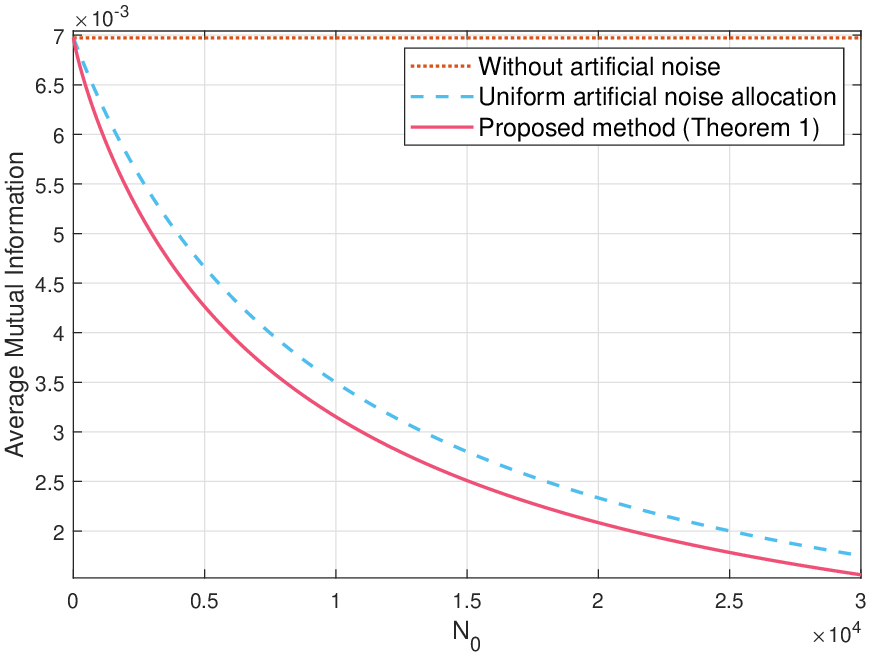}}
        \label{Fig: Gaussian_moderate_snr}
        \hfill
        
        \caption{Comparison of mutual information under the proposed method and uniform artificial noise allocation with $X_i \sim \mathcal{N}(0, \mathsf{P}_i)$ and  $\mathsf{P}_i\sim\mathcal{N}(1,0.5^2)$, where $\mathsf{P}_i$ is truncated to ensure nonnegativity: (a) $\mathsf{Z}_i=1000$, (b) $\mathsf{Z}_i=100$. 
        }
    	\vspace{-4mm}
        \label{fig:Gaussian_distribution_optimal}
    \end{figure}

     \begin{figure}[t]
        \centering
        \subfloat[]{\includegraphics[width=0.33\textwidth]{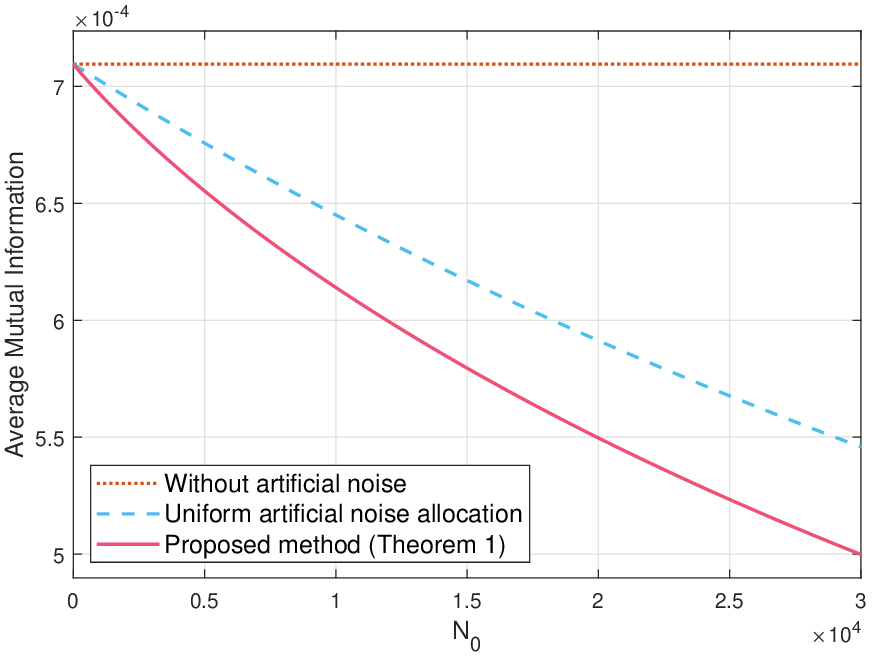}}
        \label{fig:Uniform_low_snr}
        \hfill
        \subfloat[]{\includegraphics[width=0.33\textwidth]{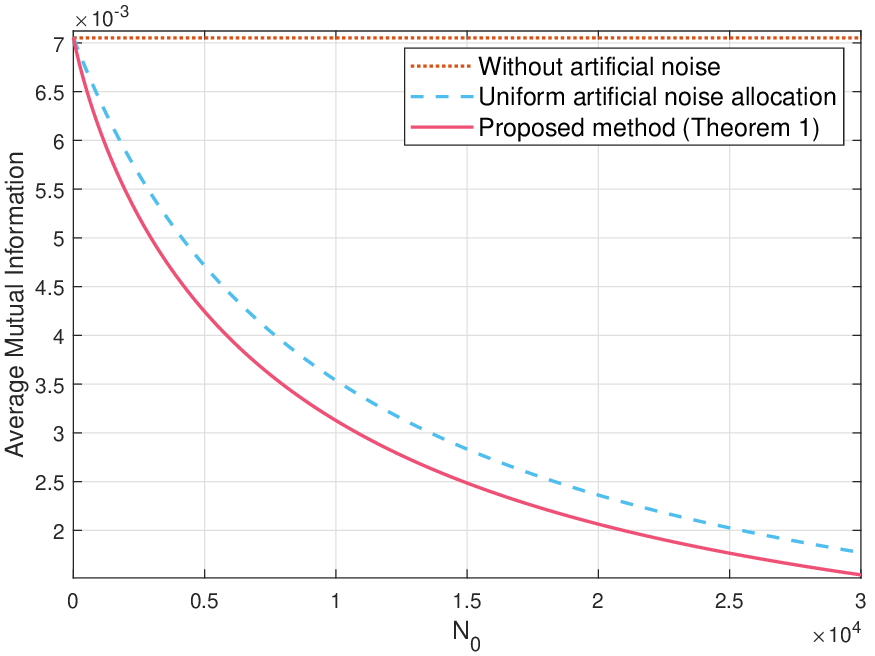}}
        \label{fig:Uniform_moderate_snr}
        \hfill
        \caption{Comparison of average (total) mutual information under the proposed method and uniform artificial noise allocation with $X_i \sim \mathcal{N}(0, \mathsf{P}_i)$ and  $\mathsf{P}_i\sim\mathcal{U}[0,2]$: (a) $\mathsf{Z}_i=1000$, (b) $\mathsf{Z}_i=100$.}
    	\vspace{-4mm}
        \label{fig:uniform_distribution_optimal}
    \end{figure}
    
    In this section, we evaluate the effectiveness of the proposed optimal artificial noise allocation.
    Specifically, we compare three approaches: (1) no artificial-noise allocation, (2) uniform artificial-noise allocation, and (3) optimized artificial-noise allocation.
    According to~\cite{das2018asni}, the average current consumption during advanced encryption standard (AES) operations is approximately \SI{1}{mA}, while effective artificial noise injection requires about \SI{17}{mA}. 
    Given that power scales with the square of current under constant resistance, this corresponds to a power ratio of approximately $1:300$. 
    Based on this, the average power consumption is normalized to $1$, and the average power of artificial noise is varied from $0$ to $300$. 
    To evaluate the impact of artificial noise, we vary the total noise power $\mathsf{N}_0 =\sum_{i=1}^m \mathsf{N}_i$ from $0$ to $300\times m$, with $m = 100$. 
    To reflect a range of practical scenarios, the input powers $\mathsf{P}_i$ are randomly drawn from either Gaussian or uniform distributions. 
        
    Fig.~\ref{fig:Gaussian_distribution_optimal} corresponds to the case where $\mathsf{P}_i$ is sampled from a Gaussian distribution, while Fig.~\ref{fig:uniform_distribution_optimal} considers a uniform distribution.
    Specifically, to achieve a \SI{50}{\%} reduction in the average mutual information, the proposed method requires \SI{19.14}{\%}, and \SI{17.92}{\%} less artificial noise power than the uniform allocation in Fig.~\ref{fig:Gaussian_distribution_optimal}(a), and (b), respectively.  
    Similarly, the proposed method reduces the artificial noise power by \SI{22.29}{\%}, and \SI{21.13}{\%} in Fig.~\ref{fig:uniform_distribution_optimal}(a), and (b), respectively.  
    These results indicate that the proposed method significantly improves noise power efficiency compared to uniform noise allocation, which is particularly critical for resource-constrained devices.
    Additional numerical results are provided in~\cite{Woo2025mutual}.
    
    \section{Conclusion}\label{sec:conclusion}
    We presented a principled information-theoretic framework for protecting cryptographic systems from power-analysis attacks through optimal artificial-noise injection. 
    Our approach aims to minimize the mutual information for arbitrary input distributions, including the Gaussian case, and we derive the corresponding optimal solution.
    Numerical results show that our proposed method reduces the mutual information more effectively than conventional uniform allocation.

\bibliographystyle{IEEEtran}
\bibliography{abrv,mybib}

@STRING{IEEE_J_CASI       = "{IEEE} Trans. Circuits Syst. {I}"}

@STRING{IEEE_J_IT         = "{IEEE} Trans. Inf. Theory"}

@STRING{JASA		  = "J. Am. Stat. Assoc."}

@STRING{AS			  = "Ann. Stat."}

@STRING{Nature		  = "Nature"}

@STRING{RSA			  = "Random Struct. Algorithms"}

@STRING{JM			  = "J. Mark."}

@STRING{DSS			  = "Decis. Support Syst."}

@STRING{CHES		  = "Proc. Int. Workshop Cryptograph. Hardw. Embedded Syst. (CHES)"}

@STRING{CRYPTO		  = "Proc. Int. Cryptol. Conf. (CRYPTO)"}

@STRING{HOST	      = "Proc. IEEE Int. Symp. Hardware Oriented Secur. (HOST)"}

@STRING{ICCAD		  = "Proc. {IEEE/ACM} Int. Conf. Comput.-Aided Design (ICCAD)"}

@STRING{ICCPS		  = "Proc. {IEEE/ACM} Int. Conf. Cyber-Phys. Syst. (ICCPS)"}

@STRING{ITA		  	  = "Proc. Inf. Theory Appl. Workshop (ITA)"}

@book{Apostol1991calculus,
  title={Calculus, Volume 1},
  author={Apostol, Tom M},
  year={1991},
  publisher={John Wiley \& Sons}
}

@inproceedings{Al2016acoustic,
    title={Acoustic side-channel attacks on additive manufacturing systems},
    author={Al Faruque, Mohammad Abdullah and Chhetri, Sujit Rokka and Canedo, Arquimedes and Wan, Jiang},
    booktitle=ICCPS,
    pages={1--10},
    year={2016},
    month = apr,
}

@inproceedings{Arsath2020param,
  author={Arsath K F, Muhammad and Ganesan, Vinod and Bodduna, Rahul and Rebeiro, Chester},
  booktitle= HOST,
  title={{PARAM: A microprocessor hardened for power side-channel attack resistance}}, 
  year={2020},
  month = dec,
  pages={23-34},
  }

@inproceedings{Brier2004correlation,
  title={Correlation power analysis with a leakage model},
  author={Brier, Eric and Clavier, Christophe and Olivier, Francis},
  booktitle=CHES,
  pages={16--29},
  month = aug,
  year={2004},
}

@book{Cover2006,
	author = {Cover, Thomas M and Thomas, Joy A},
	publisher = {Wiley-Interscience},
	address = {Hoboken, NJ},
	title = {Elements of Information Theory},
	edition = "Second",
    month = JUL,
	year = {2006}
}

@article{Das2018asni,
  title={{ASNI}: Attenuated signature noise injection for low-overhead power side-channel attack immunity},
  author={Das, Debayan and Maity, Shovan and Nasir, Saad Bin and Ghosh, Santosh and Raychowdhury, Arijit and Sen, Shreyas},
  journal=IEEE_J_CASI,
  volume={65},
  number={10},
  pages={3300--3311},
  year={2018},
  month = oct,
  publisher={IEEE}
}

@article{Efron2011tweedie,
    title={Tweedie’s formula and selection bias},
    author={Efron, Bradley},
    journal=JASA,
    volume={106},
    number={496},
    pages={1602--1614},
    year={2011},
    month = mar
}

@article{Gan2024classic,
  title={{Classic McEliece hardware implementation with enhanced side-channel and fault resistance}},
  author={Gan, Peizhou and Ravi, Prasanna and Raj, Kamal and Baksi, Anubhab and Chattopadhyay, Anupam},
  journal={Cryptology {ePrint} Archive, Paper 2024/1828},
  year={2024},
  month=nov
}

@inproceedings{Gandolfi2001electromagnetic,
  title={Electromagnetic analysis: Concrete results},
  author={Gandolfi, Karine and Mourtel, Christophe and Olivier, Francis},
  booktitle = CHES,
  pages={251--261},
  month = may,
  year={2001},
}

@inproceedings{Gattu2020power,
  title={Power side channel attack analysis and detection},
  author={Gattu, Navyata and Khan, Mohammad Nasim Imtiaz and De, Asmit and Ghosh, Swaroop},
  booktitle=ICCAD,
  pages={1--7},
  month = nov,
  year={2020}
}

@inproceedings{Gierlichs2008mutual,
  title={Mutual information analysis: A generic side-channel distinguisher},
  author={Gierlichs, Benedikt and Batina, Lejla and Tuyls, Pim and Preneel, Bart},
  booktitle = CHES,
  pages={426--442},
  year={2008},
  month = aug,
}

@article{Grushka1972characterization,
  title={{Characterization of exponentially modified Gaussian peaks in chromatography}},
  author={Grushka, Eli},
  journal={Anal. Chem.},
  volume={44},
  number={11},
  pages={1733--1738},
  year={1972},
  month = sep,
  publisher={ACS Publications}
}

@inproceedings{Guneysu2011generic,
  title={Generic side-channel countermeasures for reconfigurable devices},
  author={G{\"u}neysu, Tim and Moradi, Amir},
  booktitle=CHES,
  pages={33--48},
  year={2011},
  month = sep
}

@article{Guo2011mmse,
  title={{Estimation in Gaussian noise: properties of the minimum mean-square error}},
  author={Guo, Dongning and Wu, Yihong and Shitz, Shlomo S. and Verdú, Sergio},
  journal=IEEE_J_IT,
  volume={57},
  number={4},
  pages={2371--2385},
  year={2011},
  month = mar,
  publisher={IEEE}
}

@article{Guo2005mutual,
  title={{Mutual information and minimum mean-square error in Gaussian channels}},
  author={Guo, Dongning and Shamai, Shlomo and Verdú, Sergio},
  journal=IEEE_J_IT,
  volume={51},
  number={4},
  pages={1261--1282},
  year={2005},
  month = apr,
  publisher={IEEE}
}

@article{Issa2020maximal,
  author={Issa, Ibrahim and Wagner, Aaron B. and Kamath, Sudeep},
  journal=IEEE_J_IT, 
  month = may,
  title={{An operational approach to information leakage}}, 
  year={2020},
  volume={66},
  number={3},
  pages={1625-1657},
}

@article{Gordon1941values,
  title={{Values of Mills' ratio of area to bounding ordinate and of the normal probability integral for large values of the argument}},
  author={Gordon, Robert D},
  journal={The Annals of Mathematical Statistics},
  volume={12},
  number={3},
  pages={364--366},
  year={1941},
  publisher={JSTOR}
}

@book{Jacobs1965principles,
    title={{Principles of Communication Engineering}},
    author={Jacobs, Irwin Mark and Wozencraft, JM},
    year={1965},
    publisher = {New York: Wiley}
}

@INPROCEEDINGS{Jin2022optimal,
  author={Jin, Shan and Xu, Minghua and Bettati, Riccardo and Christodorescu, Mihai},
    booktitle = {Proc. IEEE Int. Workshop Inf. Forensics Security (WIFS)}, 
  title={Optimal Energy Efficient Design of Artificial Noise to Prevent Side-Channel Attacks}, 
  year={2022},
  month = dec,
  pages={1-6},
  }

@inproceedings{Kocher1999differential,
  title={Differential power analysis},
  author={Kocher, Paul and Jaffe, Joshua and Jun, Benjamin},
  booktitle=CRYPTO,
  pages={388--397},
  year={1999},
  month = aug,
}

@inproceedings{Kocher1996timing,
  title={{Timing attacks on implementations of Diffie-Hellman, RSA, DSS, and other systems}},
  author={Kocher, Paul C},
  booktitle=CRYPTO,
  pages={104--113},
  year={1996},
  month= aug,
}

@article{Lozano2006optimum,
	author = {Lozano, A and Tulino, A M and Verd{\'{u}}, S},
	doi = {10.1109/TIT.2006.876220},
	journal = IEEE_J_IT,
	month = jul,
	number = {7},
	pages = {3033--3051},
	title = {{Optimum power allocation for parallel Gaussian channels with arbitrary input distributions}},
	volume = {52},
	year = {2006}
}

@inproceedings{Prouff2013masking,
  title={Masking against side-channel attacks: A formal security proof},
  author={Prouff, Emmanuel and Rivain, Matthieu},
  booktitle= {Proc. Annu. Int. Conf. Theory Appl. Cryptograph.
Techn.},
  pages={142--159},
  year={2013},
  month = may,
}

@inproceedings{Shamir2000protecting,
  title={Protecting smart cards from passive power analysis with detached power supplies},
  author={Shamir, Adi},
  booktitle= CHES,
  pages={71--77},
  year={2000},
  month = aug,
}

@article{Sibson1969information,
    title     = "Information radius",
    author    = "Sibson, Robin",
    journal   = "Z. Wahrscheinlichkeitstheorie Verw. Geb.",
    publisher = "Springer Nature",
    volume    =  14,
    number    =  2,
    month = jun,
    pages     = "149--160",
    year      =  1969
}

@article{Socha2022comprehensive,
    title={A comprehensive survey on the non-invasive passive side-channel analysis},
    author={Socha, Petr and Mi{\v{s}}kovsk{\`y}, Vojt{\v{e}}ch and Novotn{\`y}, Martin},
    journal={Sensors},
    volume={22},
    number={21},
    pages={1--37},
    month = oct,
    year={2022}
}

@INPROCEEDINGS{Verdu2015alpha,
  title     = "$\alpha$-mutual information",
  author    = "Verdú, Sergio",
  booktitle = ITA,
  pages     = "1--6",
  month     =  feb,
  year      =  {2015}
}

@inproceedings{Veyrat2012shuffling,
    title={Shuffling against side-channel attacks: A comprehensive study with cautionary note},
    author={Veyrat-Charvillon, Nicolas and Medwed, Marcel and Kerckhof, St{\'e}phanie and Standaert, Fran{\c{c}}ois-Xavier},
    booktitle={Proc. Int. Conf. Theory Appl. Cryptol. Inf. Security (ASIACRYPT)},
    pages={740--757},
    month = dec,
    year={2012},
}

@article{Woo2025mutual,
  title={Mutual Information Minimization for Side-Channel Attack Resistance via Optimal Noise Injection},
  author={Woo, Jiheon and Seo, Daewon and Kim, Young-Sik and Lee, Namyoon and Cassuto, Yuval and Kim, Yongjune},
  journal={arXiv preprint arXiv:2504.20556},
  year={2025},
 url = {https://arxiv.org/abs/2504.20556},
}

@article{Wu2020case,
  title={A Case for Maximal Leakage as a Side Channel Leakage Metric},
  author={Wu, Benjamin and Wagner, Aaron B and Suh, G Edward},
  journal={arXiv:2004.08035},
  month = apr,
  year={2020}
}

\clearpage

\end{document}